# CONTRIBUTION TO THE MODELLING OF THE CORRUGATED CARDBOARD BEHAVIOUR

S. Allaoui[*], Z. Aboura[**] and M.L. Benzeggagh[**]

* Institut PRISME/LMSP, UPRES E.A. 4229, Université d'Orléans, Polytech'Orléans, 8 rue Leonard de Vinci 45072 Orléans Cedex2, France.
Mel : samir.allaoui@univ-orleans.fr .fax : 0033(0)238417329
** LRM, UMR6253, UTC, Centre de Recherche Royallieu B.P 20529 F- 60205 Compiègne Cedex, France

## ABSTRACT

The current paper summarizes studies undertaken on the corrugated cardboard. In these studies, a new approach considering the corrugated cardboard as an orthotropic material is developed. This approach permits after homogenization to simplify the numerical calculations and then use a 2D meshing of the corrugated cardboard, instead of a 3D meshing. This will makes it possible to obtain less heavy and less expensive numerical dimensioning studies. The first stage of the studies was a determination of the behaviours of the sandwich and its constituents, which passes by the definition of experimental protocols. Thereafter, an analytical model was proposed and permits to predict the homogenized behaviour of the corrugated cardboard. Finally a tool of decision-making aid was proposed.

## 1. INTRODUCTION

Corrugated cardboard pertaining to papers family is one of the most used packing currently. This success is due to the various virtues of this material: good protection of the product, low cost and can be recycled as well as biodegradable. Biodegradable product are more and more demanding as being a major concern again the protection of the environment and hence respect a durable development. In addition, the new European provisions concerning environmental protection impose a reduction of weight as well as quotas of recyclability of various packing materials. Actually, to optimize the packing (cases), the manufacturer proceed with their knowledge. They carried out mechanical tests on the final product. This causes a raw materials loss, and requires a considerable investment in human and materials means. The possibility of acting upstream is desirable and would make it possible to develop a rigorous and cheaper method. This demarche needs a tool of decision-making aid using simple model.

The corrugated cardboard is an orthotropic sandwich with the surface plies (facing) providing bending stiffness, separated by a lightweight bending core (fluting) that provides shear stiffness. Two main directions characterise this material. The first noted MD (machine direction) corresponds to the direction of manufacturing of the material. It coincides with the "x" axis (Figure 1). The second noted CD (cross direction) corresponds to the transverse direction and coincides with the "y" axis. To refer to the out-of-plane direction (through-thickness), a third direction ZD is introduced as depicted by figure 1. It is generally composed by three paperboard constituents: upper layer, lower layer and fluting. We observe the same directions for the paperboard, where the machine direction corresponds to the feel orientation of the cellulose fibres. This privileged orientation is due to the continuous nature of the material process.

The Approach used in different studies, consider the corrugated cardboard as structure. These studies used norms relative to the composite or metallic sandwich structure to apprehend mechanical behaviour of the corrugated cardboard. The main tests are three or four points bending (ASTM C393-62), as well as shear tests (ASTM C273-61). Norstrand et al. [1] have evaluated the transverse shear stiffness of corrugated cardboard by ASTM block shear test and by three-point bending test. Nevertheless, this approach does not allow determining the stiffness matrix of such a material and makes therefore its homogenization for a finite elements calculation impossible. It is obvious that the possibility exists always to mesh the structure completely (skins and core) [2–4] leading to an extremely expensive process.

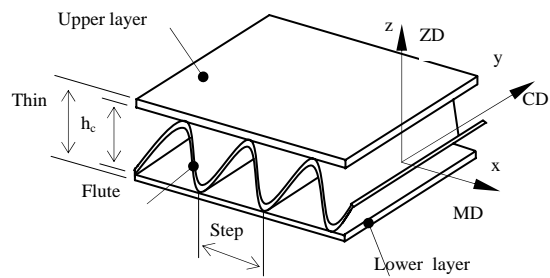

Figure 1: Cardboard panel geometry

In this study, another alternative being well developed consists to replacing the corrugated cardboard by an equivalent orthotropic layer [5]. This approach will permit after homogenization to simplify the numerical calculations and then use a 2D meshing of the corrugated cardboard (figure 2), instead of a 3D meshing extensively used in the literature. This will makes it possible to obtain less heavy and less expensive numerical dimensioning studies of cases for example.

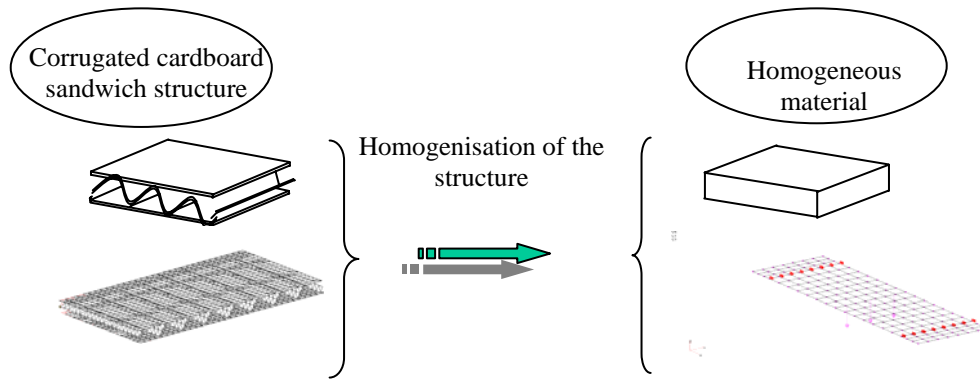

Figure 2: Homogenization approach

This approach is articulating around two points:
• The first point is to dedicate the different experimental protocols finalized for the sake of determination the mechanical behaviour of the constituents (skins and core) as well as those of the corrugated cardboard. The main difficulty resides in the experimental precautions in order to identify the necessary parameters in the absence of the pre-workbenches protocols.
• The second point is dedicate to the analytical modelling. Thus, by using the mechanical properties and geometrical parameters of the constituents (step of fluting, thickness of skins and core), this model should be able to predict the homogenized behaviour of the corrugated cardboard.

A comparison between the experimental observations, analytic model and finite elements prediction allows judging the performances of the proposed model.

## 2. EXPERIMENTAL TESTS

The first part of the study stuck to the establishment of specific experimental procedures tests that considering the corrugated cardboard like a material and not a structure. They are extending to the case of the paperboard in order to obtain their behaviour, which will be used in the modelling method. These procedures are inspired by the norms relative to the composite and metallic sandwich structure [5]. One of the significant points in these procedures is the rigidification of the specimen heads by injecting a polyester resin between fluting and cardboard skins to fill the existing emptiness or by impregnating the extremities in the resin for the paperboard specimens. Indeed, this permit to avoid the bruising of the specimen heads at the griping time and thus the use of work-holding device. In the other hand, the instrumentation required a particular attention because the material becomes sensitive to any handling during the mechanical tests. Several means were used and the optical method using marks tracking technique was validated because it is the most adapted for tests at different atmospheric conditions [6].

After having established the various protocols, mechanical tests were carried out on corrugated cardboard and the paperboard. Initially tensile, twisting, shear and three point bending tests were carried out in order to determine the elastic and failure parameters of the paperboard and the sandwich. Figure 3 represents some examples of the corrugated cardboard behaviour under uni-axial solicitations. Generally, we have noted that the behaviour in the cross direction and the machine direction are different. For each material and in each direction, we have determined the mechanical parameters but also the chronology of these behaviours.

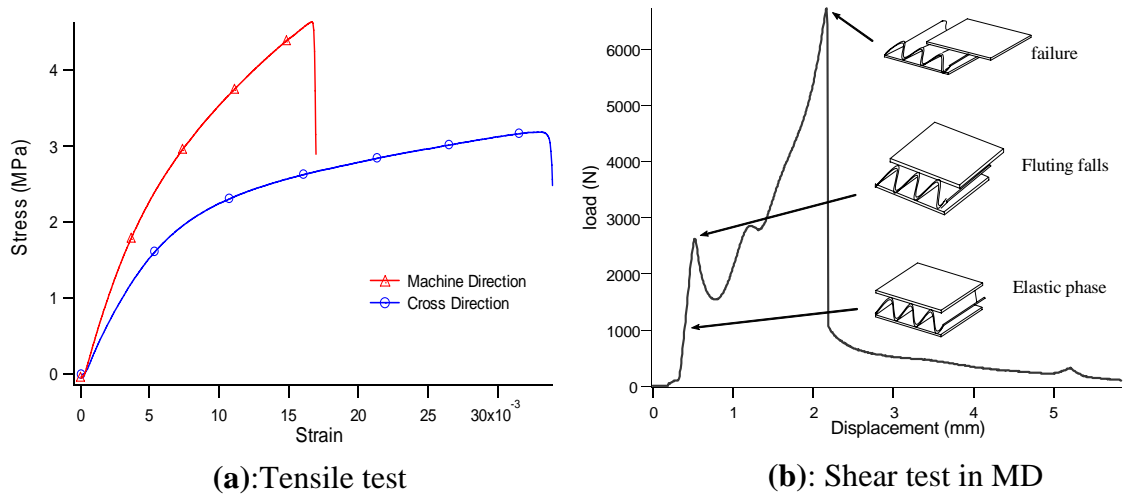

**(a)**: Tensile test    **(b)**: Shear test in MD

Figure 3: Examples of the corrugated cardboard under uni-axial solicitations

In the case of the tensile test, table 1 illustrates elastic and failure properties obtained. We note that these mechanical parameters are twice more significant in the machine direction than in the cross direction, which correspond to the celluloses fibres orientation and the flute one. Concerning the corrugated cardboard, it appears that the load failure, in MD, is equal to the sum of the maximum failure loadings of the two main layers. On the other hand, in CD, it is equal to the sum of those of the three constituent's failures loads (layers and flute). This illustrates the principal teaching of the different tests, where the corrugated cardboard behaviour is governed by the layers in the machine direction and by all the constituents in the cross direction.

|               | Material             |             |             |         |
| ------------- | -------------------- | ----------- | ----------- | ------- |
| Direction     | Corrugated cardboard | Upper layer | Lower layer | Flute   |
| **Young Moduli (MPa)** | | | | |
| **MD**        | 644.45               | 4322.20     | 4433.48     | 5106.67 |
| **CD**        | 433.10               | 2048.27     | 2032.91     | 1962.35 |
| **Load (N)**  | | | | |
| **MD**        | 285.35               | 134.83      | 146.81      | 123.75  |
| **CD**        | 185.79               | 60.71       | 61.87       | 43.58   |

Table 1: elastic and failure parameters of the corrugated cardboard and its constituents

As we can see on the Characteristic stress–strain curves (figure 3a), mechanical behaviour of the corrugated cardboard (in the two in-plane directions) is composed by linear part followed-up by non-linear part. We note that the same behaviour phases were found in the case of the paperboard. The non-linearity is accentuated in the cross direction. This phenomenon depends on the cellulose fibres, the hydrogen bonds and the rate of the moisture content. This more marked non-linearity in the cross direction can be due to the fact that the paperboard mechanical behaviour, therefore the sandwich one, are more governed by hydrogen bonds that are more requested in CD than the cellulose fibres.

To highlight this viscous behaviour, uni-axial tensile tests at different strain rate ($\dot{\varepsilon}$ going from $6 \times 10^{-5}$ s$^{-1}$ to $12 \times 10^{-3}$ s$^{-1}$) and other with strain recoveries were carried out. These tests make us able to identify that the elastic and failure parameters evolve more in CD with the strain rate solicitation [7]. This evolution is more significant in the case of the layers which have a more significant coating. This coating, constituted by a polymer material, is a surface treatment used to improve the printing quality but also to have a better surface aspect. So, the coating undoubtedly interferes on the material behaviour.

Loading–unloading-holding tensile tests carried out on the skins, in the two directions, revealed a residual plastic strain. Thus, non-linearity of the paperboard is governed by a viscoplastic phenomenon. The slopes of unloading shows a curved shape, which indicate the presence of energy dissipation, so a loss of stiffness. This dissipation is due to the presence of a damage coupled with the viscoplastic phenomenon. This damage has been highlighted by cyclic tests on the paperboard and the sandwich.

After that, relaxation tensile tests under SEM were carried out in order to identify the coupled damage-viscous-plasticity. During these tests carried out on micro-specimens of the paperboard, fractographies were taken to show the mechanisms which govern this damage. Mechanical parameters obtained by theses tests were not exploited because of the scale effect. Only physical phenomena have been considered. These tests confirm the presence of viscoelastic phenomenon at the elastic phase where we note a stress relaxation with the first strain holding (figure 4). This phenomenon is more significant in the cross direction than in the machine direction. This is due to the fact that the hydrogen bonds between fibres, which can have a viscous behaviour, are more requested in CD. These observations confirm those make in the precedent static tests. The damage intervenes at the linear part of the behaviour by the locally appearance of microscopic cracks. By increasing load level, microscopic cracks appear in greater number before starting of macroscopic crack, which causes the final failure of material (figure 4). The damage of the paperboard intervenes in several forms. We note fibre skirting, fibres failures or even coating damage. These mechanisms are found in the two in-plane directions (MD and CD), but with more or less significant proportions [7].

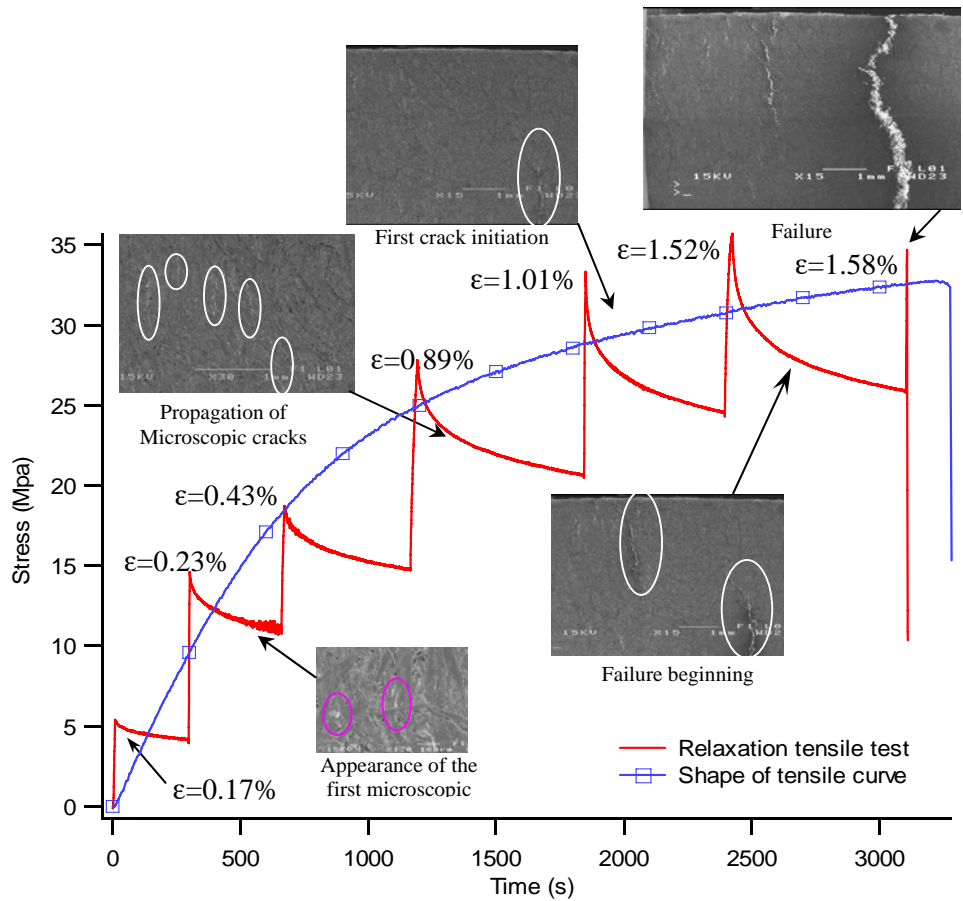

Figure 4: Cross direction relaxation tensile test under SEM of the paperboard.

Mechanical tests carried on the corrugated cardboard and its constituents allowed us to identify their elastic and failure parameters. In addition, we have highlighted the complexity of the material behaviour, which is governed by a coupled phenomenon (viscous-plastic-damage). However, these behaviours evolve with the rate relative humidity, phenomenon in which we were interested [6]. Indeed, we have noted a loss of performance of more than 50% while passing from 50%RH to 90RH. The whole of the obtained experimental results constitutes an essential database to the development of an analytical model, which will be able to predict the mechanical behaviour of the sandwich.

### 3. ANALYTICAL MODELLING

An analytical model was developed. It is based on the classical laminate theory was feed by mechanical and geometrical characteristics of the sandwich constituents. In the first time, the model was applied to predict elastic behaviour of the sandwich.

**Formulation of the analytical model**

To analyse the elastic behaviour of the corrugated cardboard, the proposed approach is based on the classical laminate theory. The model is inspired from Ishikawa et al [8], Aboura [9] and Scida et al. [10] related to the elastic behaviour modelling of woven composites materials. A unit cell representative of the corrugated cardboard is defined as in figure 5 and considered the skins and the undulated fluting as an assembling of many infinitesimal elements dx of unidirectional lamina oriented at different angles. The classical laminate theory is then applied to each element.

$$\left\{ \begin{array}{c} N_i \\ M_i \end{array} \right\} = \left[ \begin{array}{cc} A_{ij} & B_{ij} \\ B_{ij} & D_{ij} \end{array} \right] \left\{ \begin{array}{c} \varepsilon_j \\ k_j \end{array} \right\} \quad (i,j = 1,2 \text{ and } 6) \tag{1}$$

in which $A_{ij}$, $B_{ij}$ and $D_{ij}$ are the in-plane stiffness for each infinitesimal element dx.

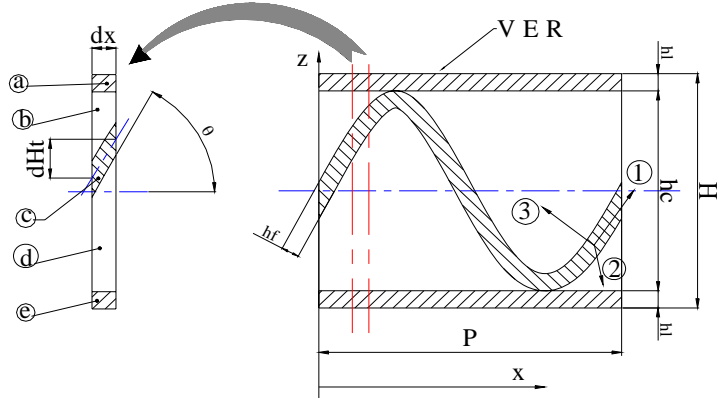

Figure 5: Unit cell representative (VER) of the corrugated cardboard

These are defined by:

$$(A, B, D) = \int_{-h/2}^{h/2} (1, z, z^2) Q_{ij} dz \tag{2}$$

in which

$$Q_{ij} = \left[ \begin{array}{ccc} \dfrac{E_x}{1 - \nu_{xy} \nu_{yx}} & \dfrac{\nu_{xy} E_y}{1 - \nu_{xy} \nu_{yx}} & 0 \\ \dfrac{\nu_{yx} E_x}{1 - \nu_{xy} \nu_{yx}} & \dfrac{E_y}{1 - \nu_{xy} \nu_{yx}} & 0 \\ 0 & 0 & G_{xy} \end{array} \right] \quad (i,j = 1,2,6) \tag{3}$$

$Q_{ij}$ is evaluated for each constituent of the corrugated cardboard unit cell. This means that the superior and inferior skins and the fluting undulation are taken into account. The local stiffness of each infinitesimal element depends on the constituent elastic properties as well as on the fluting orientation defined by the local off-axis angle $\theta(x)$. This angle is calculated from the fluting median fibre function $Ht(x)$:

$$\theta(x) = \tan^{-1}\left(\frac{dHt(x)}{dx}\right) \tag{4}$$

$Ht(x)$ is assumed to be of sinusoidal form with a maximum thickness of hc:

$$Ht(x) = \frac{hc}{2} \sin\left(2\pi \frac{x}{P}\right) \tag{5}$$

Once the $Q_{ij}$ terms are calculated, the in-plane stiffness coefficients can be evaluated for each element in different regions of the unit cell by using equation (2). The global matrices $A_{global}$, $B_{global}$ and $D_{global}$ for the corrugated cardboard unit cell are calculated numerically from the local matrices A, B and D evaluated for each infinitesimal element with an average in the "x" direction by:

$$(A, B, D)_{global} = \frac{1}{P} \int_0^P A(x), B(x), D(x) dx \tag{6}$$

From the global matrices [A], [B] and [D], the effective Young's modulus, shear modulus and Poisson's ratio of the corrugated cardboard can be obtained.

**Results of homogeneous elastic properties**

Two layers and one flute compose the corrugated cardboard used for the model validation. The mechanical characteristics of the constituents are obtained by tensile tests according to experimental protocols developed in precedent studies [5, 6]. They are shown with the geometrical characteristics in the table 2.

|  | h (mm) | $E_{MD}$ (Mpa) | $E_{CD}$ (Mpa) | $\nu_{xy}$ | $\nu_{yx}$ |
|---|---|---|---|---|---|
| Upper layer | 0.235 | 4514.53 | 1895.83 | 0.282 | 0.215 |
| Flute | 0.235 | 4703.75 | 1854.3 | 0.353 | 0.088 |
| Lower layer | 0.19 | 4458.6 | 1944.5 | 0.277 | 0.115 |

Table 2: Mechanical and geometrical characteristics of the sandwich constituents

After that, the model was used to evaluate the homogeneous characteristics of the sandwich. Table 3 shows the results of the homogenization. We note that the model results are in good agreement with the experience. All the analytical homogeneous characteristics are given with an error less than 9% compared to the experimental ones. This error is due to the effect of the corrugated cardboard manufacture process and the fact that the Unit cell representative (VER) doesn't take account of the interface between the skins and the fluting.

|  | $E_{MD}$ (Mpa) | $E_{CD}$ (Mpa) | $\nu_{xy}$ | $\nu_{yx}$ |
|---|---|---|---|---|
| Experimental characteristics | 656 | 412.38 | 0.251 | 0.125 |
| Model results | 711.99 | 391.47 | 0.246 | 0.135 |
| Model / Exp. (%) | 8.53 | -5.07 | -1.87 | 8.33 |
| Model results after correction | 676.1 | 390.78 | 0.246 | 0.142 |
| Model after correction / Exp. (%) | 3.06 | 5.24 | -1.87 | 13.88 |

Table 3: Comparison between experimental and model results

To evaluate the effect of the process, we determined the mechanical characteristics of the skins before and after manufactured of the corrugated cardboard. We carried out micro tensile tests on skins gives by the manufacturer and skins extracted on the structure. Tests were carried out in the cross and machine direction. We found that the manufactured process does not affect the characteristics of the lower layer and the flute, when the upper layer loses 13% of its characteristics in the machine direction. After that we have use the news mechanics characteristics of the upper layer as inputs for the analytical model. We note that the error in estimation of the Young modulus in the machine direction passes from 8.53% to 3.06%. In the same time the other characteristics do not evolve significantly

**Modelling of the inelastic behaviour**

The analytical model developed was carried out in the elastic range and permit to predict Homogeneous mechanical characteristics in good agreement with those obtained by experimental tests. Beyond this elastic part, the corrugated cardboard has non-linear behaviour. The extension of the analytical model to the prediction of the behaviour beyond elasticity requires the knowledge of the behaviour laws of the corrugated cardboard components. In this approach, the model will be fed by experimental results resulting from tests carried out on the layers and the flute. The model will be extending to the prediction of non-linear uniaxial tensile behaviour of the material. This extension needs to know the stress-strain curve in MD and CD of the sandwich constituents. Assuming strain continuity, the local stress is computed in each element of the unit cell at every increment of macroscopic stress by using the stress-strain curves of the skins and fluting. A new stiffness is then determined. By homogenisation on all unit cell, the global stiffness can be calculated and then the prediction of non-linear behaviour of the corrugated cardboard.

Figure 6 shows a good agreement between the behaviour simulated by the model in the machine direction and the experimental results one. On the other hand, the agreement is worse in the cross direction (figure 6b). Indeed after having approached elastic phase correctly, the model over-estimates the inelastic phase. This phenomenon is due to the effect of the interface between the two skins and the flute. This interface intervenes when the material is requested mechanically. It is made up of starch, which is used as adhesive between the tops of the flute and the skins. During the tensile test, the request of the interface causes a local damage due to the high rigidity of the starch compared to the paperboard. This phenomenon does not interfere in the machine direction.

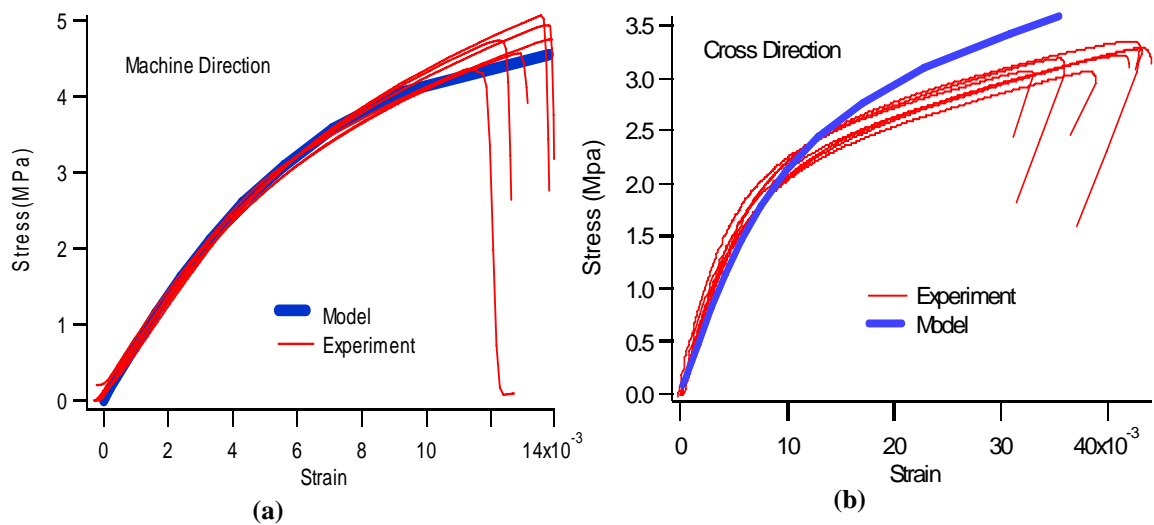

Figure6: Comparison between experimental and model results of the uni-axial tensile behaviour.

Modelling of the global behaviour of the sandwich was then used to predict the behaviour of the sandwich at different atmospherics conditions. The results obtained are in good agreement with the experimental ones. However, a parametrical study was done in order to evaluate the geometrical parameters (like layers thickness, core thickness, fluting step …etc.), which influence more the sandwich behaviour.

## 4 NUMERICAL SIMULATION

In order to assess the relevance of the homogenization method developed for this purpose, a finite element analysis was used using the homogeneous mechanical parameters obtained by the model. The main goal is to verify the possibility to use a 2D meshing of the corrugated cardboard, instead of a 3D meshing extensively used in the literature. A finite element models were created for mechanical tests such as the three points bending and free vibration tests. The results of the three points bending with an applied load of 5N are presented in this study. The geometrical characteristics of this model are chosen similar to those used in the corresponding experiments. The used element is the basic triangular 3-nodes flat shell. Table 4 presents a comparison between experimental and the two finite element approaches results. The agreement between experimental and the two FE approaches is correct. We note that the bending stiffness are estimate with a maximum error of 12% the cross direction. For the 3D meshing the stiffness is estimate with an error of 7.86% in the cross direction. However, for an overall behaviour, the time calculation of the 2D approach is 10 times lower than that of the 3D approach.

|  | 2D Approach | 3D Approach | Experiment |
|---|---|---|---|
| Bending rigidity (N/mm) | 2.017 | 2.47 | 2,29±0,09 |
| CPU Time (s) | 0.3 | 3 | / |

Applied load: 5 N.

Table 4: Comparison between FE and experimental results

It appears that the simplified homogenisation procedure is adequately accurate and fast for effectively analysing corrugated cardboard panel in the preliminary and optimisation design stages. For this aim, a tool of decision-making aid using the results obtained by this various work is developed. This tool makes it possible to predict the homogenized mechanical parameters of the corrugated cardboard, but also to choose the adequate constituents (layers and flute) in order to avoid over sizing the packing (case for example).

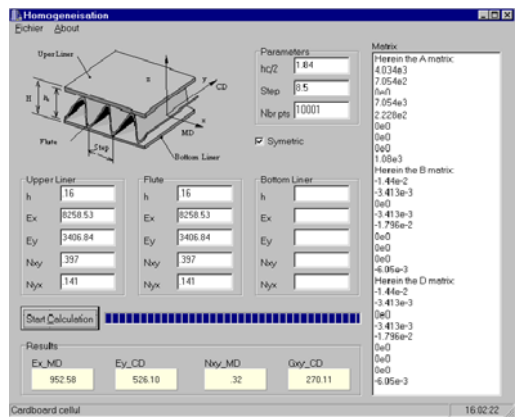

Figure 7: Interface of the tool of decision-making aid

## 5. CONCLUSIONS

This work summarizes different studies devoted to the study of the mechanical behaviour of the corrugated cardboard. The approach adopted considers the sandwich as an orthotropic material and not a structure. Different experimental protocols were developed in order to characterize the homogenized behaviour of the sandwich. The tests carried out makes it possible to better apprehend this behaviour which is more complex than as described in the literature. Thus we have highlighted a coexistence of viscous, plastic and damage phenomenon.

After that, an analytical model has been proposed. It permits to predict the homogeneous mechanic parameters in a good agreement with those obtained with the experimental tests. This model have been extend to the global behaviour of the sandwich and gives satisfaction.

In order to assess the relevance of the homogenisation method developed for this purpose, a finite element analysis is used. The main goal of this work is to verify the possibility to use a 2D meshing of the corrugated cardboard, instead of a 3D meshing. It was found that the simplified homogenisation procedure is adequately accurate and fast for effectively analysing corrugated cardboard panel in the preliminary and optimum design stages. A tool of decision-making aid was then proposed in order to help the engineers of engineering department, in the choice of the corrugated cardboard constituents. This makes it possible to not oversize the final product (like case) and thus a raw material profit.